\documentclass[pdflatex,twocolumn,showpacs,prl,superscriptaddress,amsmath,amssymb]{revtex4-1}

\usepackage{multirow}
\usepackage{graphicx}
\usepackage{array}
\usepackage[dvips]{epsfig}
\begin{document}

\preprint{APS/123-QED}

\title{Reduced quasifission competition in fusion reactions forming neutron-rich heavy elements}

\author{K. Hammerton}
\email{hammerto@nscl.msu.edu}
\affiliation{National Superconducting Cyclotron Laboratory, Michigan State University, East Lansing, Michigan 48824, USA}
\affiliation{Department of Chemistry, Michigan State University, East Lansing, Michigan 48824, USA}
\author{Z. Kohley}
\affiliation{National Superconducting Cyclotron Laboratory, Michigan State University, East Lansing, Michigan 48824, USA}
\affiliation{Department of Chemistry, Michigan State University, East Lansing, Michigan 48824, USA}
\author{D.~J. Hinde}
\affiliation{Department of Nuclear Physics, Research School of Physics and Engineering, Australian National University, Canberra, Australian Capital Territory 2601, Australia}
\author{M. Dasgupta}
\affiliation{Department of Nuclear Physics, Research School of Physics and Engineering, Australian National University, Canberra, Australian Capital Territory 2601, Australia}
\author{A. Wakhle}
\affiliation{National Superconducting Cyclotron Laboratory, Michigan State University, East Lansing, Michigan 48824, USA}
\affiliation{Department of Nuclear Physics, Research School of Physics and Engineering, Australian National University, Canberra, Australian Capital Territory 2601, Australia}
\author{E. Williams}
\affiliation{Department of Nuclear Physics, Research School of Physics and Engineering, Australian National University, Canberra, Australian Capital Territory 2601, Australia}
\author{V.~E. Oberacker}
\affiliation{Department of Physics and Astronomy, Vanderbilt University, Nashville, Tennessee 37235, USA}
\author{A.~S. Umar}
\affiliation{Department of Physics and Astronomy, Vanderbilt University, Nashville, Tennessee 37235, USA}
\author{I.~P. Carter}
\affiliation{Department of Nuclear Physics, Research School of Physics and Engineering, Australian National University, Canberra, Australian Capital Territory 2601, Australia}
\author{K.~J. Cook}
\affiliation{Department of Nuclear Physics, Research School of Physics and Engineering, Australian National University, Canberra, Australian Capital Territory 2601, Australia}
\author{J. Greene}
\affiliation{Physics Division, Argonne National Laboratory, Lemont, Illinois 60473, USA}
\author{D.~Y. Jeung}
\affiliation{Department of Nuclear Physics, Research School of Physics and Engineering, Australian National University, Canberra, Australian Capital Territory 2601, Australia}
\author{D.~H. Luong}
\affiliation{Department of Nuclear Physics, Research School of Physics and Engineering, Australian National University, Canberra, Australian Capital Territory 2601, Australia}
\author{S.~D. McNeil}
\affiliation{Department of Nuclear Physics, Research School of Physics and Engineering, Australian National University, Canberra, Australian Capital Territory 2601, Australia}
\author{C.~S. Palshetkar}
\affiliation{Department of Nuclear Physics, Research School of Physics and Engineering, Australian National University, Canberra, Australian Capital Territory 2601, Australia}
\author{D.~C. Rafferty}
\affiliation{Department of Nuclear Physics, Research School of Physics and Engineering, Australian National University, Canberra, Australian Capital Territory 2601, Australia}
\author{C. Simenel}
\affiliation{Department of Nuclear Physics, Research School of Physics and Engineering, Australian National University, Canberra, Australian Capital Territory 2601, Australia}
\author{K. Stiefel}
\affiliation{National Superconducting Cyclotron Laboratory, Michigan State University, East Lansing, Michigan 48824, USA}
\affiliation{Department of Chemistry, Michigan State University, East Lansing, Michigan 48824, USA}

\date{\today}

\begin{abstract}
Measurements of mass-angle distributions (MADs) for Cr~+~W reactions, providing a wide range in the neutron-to-proton ratio of the compound system, $(N/Z)_{\mathrm{CN}}$, have allowed for the dependence of quasifission on the $(N/Z)_{\mathrm{CN}}$ to be determined in a model-independent way. Previous experimental and theoretical studies had produced conflicting conclusions.  The experimental MADs reveal an increase in contact time and mass evolution of the quasifission fragments with increasing $(N/Z)_{\mathrm{CN}}$, which is indicative of an increase in the fusion probability.  The experimental results are in agreement with microscopic time-dependent Hartree-Fock calculations of the quasifission process.  The experimental and theoretical results favor the use of the most neutron-rich projectiles and targets for the production of heavy and superheavy nuclei.
\end{abstract}


\maketitle

\par
The existence and properties of superheavy nuclei provide stringent benchmarks for theoretical predictions of the nuclear landscape and our understanding of the nuclear force.  With the production of each new superheavy element (SHE) the boundaries of the periodic table and chart of the nuclides are extended~\cite{Hamilton2013,Hofmann2000}. In recent years, this has come through the formation of new SHEs using heavy-ion fusion reactions~\cite{Hamilton2013,Ogan2012,Ogan2007}. Theory predicts that the region around $N=184$ and $Z = 114-126$ will contain
the next spherical shell closure~\cite{Bender1999}. The existence of this ``island of stability'' is supported by the observed increase in stability of superheavy nuclei with neutron numbers nearing $N = 180$~\cite{Hamilton2013, Ogan2010}.   Further exploration of this region and synthesis of new neutron-rich heavy or superheavy nuclei will require increasingly neutron-rich projectiles and targets~\cite{Hamilton2013,Ogan2012,Bender1999,Bender2001,Nazarewicz2002}. Neutron-rich beams from next-generation radioactive ion beam (RIB) facilities will offer the first opportunities to explore their use for the production of new neutron-rich ``light'' superheavy isotopes and continue to push to the limits of stability~\cite{Hofmann2001, Loveland2007, Smol10, Sar13}.

The production cross section for the superheavy evaporation residues (ER) can be defined as,
\begin{equation}
\sigma_{\mathrm{ER}}~=~\sum^{\infty}_{J=0}~\sigma_{\mathrm{cap}}(E_{\mathrm{c.m.}},J)~P_{\mathrm{CN}}(E^{*},J)~W_{\mathrm{sur}}(E^{*},J)
\label{eqER}
\end{equation}
where $\sigma_{\mathrm{cap}}$ is the cross section for the projectile and target to come together (capture) and form a dinuclear system, $P_{\mathrm{CN}}$ is the probability that the dinuclear system fuses rather than re-separates in the quasifission process~\cite{Toke1985,Back1985}, and $W_{\mathrm{sur}}$ is the probability for the fused compound nucleus to survive against fission.  While experiments have, in general, observed increased ER cross sections for systems with increased neutron-richness~\cite{Drag08,Ogan2001,Ogan13,Wang2010,Hofmann2001,Liang07}, it is imperative to disentangle the dependence of each component of Eq.~(\ref{eqER}) on the neutron excess in order to properly assess the opportunities neutron-rich beams may offer.  For example, the increased cross sections observed for SHEs nearing $N=184$~\cite{,Ogan2010,Ogan2012} have been linked to the increased compound nucleus fission barriers, and thus $W_{\mathrm{sur}}$, but $P_{\mathrm{CN}}$ may also benefit from a change in neutron-richness.


\par
The largest uncertainty in the predictions for SHE production rates is $P_{\mathrm{CN}}$, estimated to have an uncertainty of $1-2$ orders of magnitude~\cite{Loveland2007}, owing to the complex nature of the quasifission process~\cite{Toke1985,Back1985,Zag2001,Berriman2001,Yanez2013,Hamilton2013}.  Quasifission is influenced by many factors, including entrance-channel mass asymmetry [$\alpha = (A_{\mathrm{target}} - A_{\mathrm{proj}})/(A_{\mathrm{target}} + A_{\mathrm{proj}})$]~\cite{Berriman2001}, fissility of the compound nucleus~\cite{Swiatecki1982,Bjornholm1982}, magicity of the projectile/target~\cite{Itkis2007,Simenel2012}, reaction energy~\cite{Tsang83, Backetal1985}, and nuclear deformation~\cite{Hinde1995,Nish01,Lin2012}. Experimental studies have indicated that quasifission is dependent on the neutron-richness of the compound system, characterized by $(N/Z)_{\mathrm{CN}}$. However, the previous work led to contradictory conclusions, based on strongly model-dependent analyses~\cite{Simenel2012,Sahm1985,Lesko1986, Vinod2008,Liang2012,Adamian2000,Wang2010,Knyazheva2007}.  If quasifission is reduced in more neutron-rich systems this would result in increased fusion probabilities and would help compensate for the decreased beam intensities of neutron-rich radioactive ion beams.


In this Letter, the dependence of quasifission competition of the neutron-richness of the compound nucleus is analyzed via measurements of mass-angle distributions which allow for the characterization of quasifission in a model-independent way~\cite{Toke1985, Bock1982, DuRietz2011, DuRietz2013, Wakhle2014}. To minimize the sensitivity to deformation and spherical closed shells, we have carefully chosen a set of Cr~+~W reactions (Table~I) that should show the underlying dependence of the quasifission competition on the neutron-richness of the isotope being formed through fusion.  Between the most neutron-deficient ($^{50}$Cr $+ ^{180}$W) and neutron-rich ($^{54} $Cr $+ ^{186}$W) systems there is a change of $10$ neutrons.  The Cr~+~W reactions all have the same charge product, Z$_{\mathrm{p}}$Z$_{\mathrm{t}}$, of 1776 and only $^{52}$Cr has a closed shell with $N = 28$. While the W targets are deformed their deformations are relatively similar ($\beta_{2}$ from 0.254 to 0.225~\cite{Raman2001}) and should not lead to any significant differences between the reactions~\cite{Lin2012}. By choosing above-barrier beam energies with constant $E_{\mathrm{c.m.}}/V_{\mathrm{B}}$ ($E_{\mathrm{c.m.}}$ is the center-of-mass energy and $ V_{\mathrm{B}}$ is the Bass barrier~\cite{Bass1980}) all orientations are sampled.  Finally, the conclusions drawn from the experimental results are compared to microscopic time-dependent Hartree-Fock (TDHF) calculations that predict the mean observables measured in the MADs.

\begin{table}
\caption{Center-of-mass energy $E_{\mathrm{c.m.}}$, Excitation energy E$^{*}$, $N/Z$ of the compound nucleus $(N/Z)_{\mathrm{CN}}$, and Mass asymmetry $\alpha$ for each of the systems. Each reaction was measured at $E_{\mathrm{c.m.}} / V_{\mathrm{B}} = 1.13$.}
 \centering
 \setlength{\extrarowheight}{2pt}
\begin{tabular}{ccccc}
  \hline
  \hline
 System & $E_{\mathrm{c.m.}} (MeV)$ &E$^{*}$ (MeV) &  $(N/Z)_{\mathrm{CN}}$ &$\alpha$ \\
  \hline
  $^{50}$Cr $+ ^{180}$W&222.6&64.59&1.35&0.565\\
  $^{52}$Cr $+ ^{180}$W&221.2&59.07&1.37&0.552\\
  $^{54}$Cr $+ ^{180}$W&219.8&56.73&1.39&0.538\\
  $^{50}$Cr $+ ^{186}$W&221.0&71.65&1.41&0.576\\
  $^{52}$Cr $+ ^{184}$W&220.1&62.40&1.41&0.559\\
  $^{54}$Cr $+ ^{182}$W&221.0&57.57&1.41&0.542\\
  $^{54}$Cr $+ ^{184}$W&218.9&59.03&1.43&0.546\\
  $^{54}$Cr $+ ^{186}$W&218.3&60.85&1.45&0.55\\
  \hline
  \hline
\end{tabular}
\label{tableI}
\end{table}

\begin{figure*}
\centering
\includegraphics*[width=0.75\textwidth]{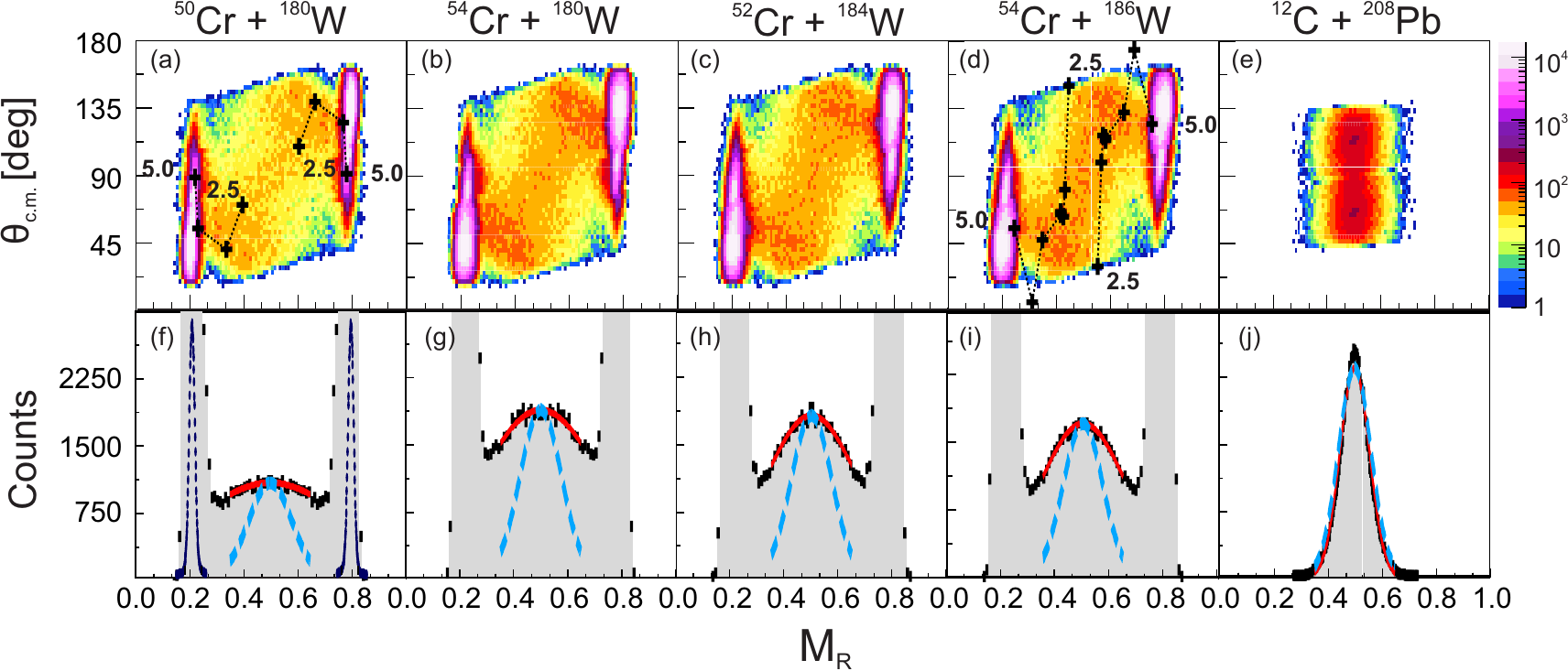}
\caption{Measured MADs for systems $^{50}$Cr $+ ^{180}$W, $^{54}$Cr$ + ^{180}$W,
$^{52}$Cr$ + ^{184}$W, and $^{54}$Cr$ + ^{186}$W are shown in panels (a) - (d), respectively. The contour scale (z-axis) of each MAD is scaled to match the maximum fission fragment yield.   The black crosses in (a) and (d) are the results of TDHF calculations at various impact parameters from $b = 2.5 - 5$~fm.  Panels (f) - (i) show the projected mass ratio distributions.  The solid red line represents a Gaussian fit over the region from $0.35 \leq M_{\mathrm{R}} \leq 0.65$.  The dashed blue line is an estimate for a pure fusion-fission mass distribution.  The mass resolution can be seen from the width of the elastic scattering peaks which are presented (scaled by 0.09) in panel (f). Panel (e) and (j) show the MAD and mass distribution for $^{12}$C~$+$~$^{208}$Pb, where no quasifission is present.  }
\end{figure*}

\par
Beams of $^{50,52,54}$Cr were provided by the 14UD electrostatic accelerator and superconducting LINAC at the Heavy Ion Accelerator Facility at the Australian National University (ANU). These beams bombarded isotopically enriched targets of $^{180,182,184,186}$W with thicknesses ranging from $43-97~\mu$g/cm$^{2}$ mounted on $40-60~\mu$g/cm$^{2}$ carbon backings. All reactions were measured at $E_{\mathrm{c.m.}} / V_{\mathrm{B}} = 1.13$. Fragments resulting from fusion-fission and quasifission (collectively termed fission-like) were detected in coincidence using the ANU CUBE detector system \cite{Hinde1996}, which consists of two large-area, position sensitive multiwire proportional counters (MWPCs), each with an active area of $28\times36$ cm$^2$~\cite{Hinde1996, DuRietz2013}.  The MWPCs covered laboratory scattering angles of $5^{\circ} < \theta < 80^{\circ}$ and $ 50^{\circ} < \theta < 125^{\circ}$. From the measured velocity vectors of the coincident fragments the mass ratio, $M_{\mathrm{R}} = m_{1} / (m_{1} + m_{2})$ where $m_1$ and $m_2$ are the masses of the fission fragments, could be determined~\cite{DuRietz2013}.

\par
The MADs, showing $\theta_{\mathrm{CM}}$ as a function of $M_{\mathrm{R}}$, for a subset of the reactions are shown in Fig. 1~(a)-(d). The additional measured MADs are shown in Ref.~\cite{SuppMat}.  The intense vertical bands around $M_{\mathrm{R}} = 0.22$ and $M_{\mathrm{R}} = 0.78$ are from elastic and quasielastic scattering of the projectile and target. In between these bands, in the region of $0.3 < M_{\mathrm{R}} < 0.7$, the fission-like fragments are observed. Previous work has shown that the MADs are a sensitive probe of the dynamics of the quasifission reactions~\cite{Toke1985, DuRietz2013, DuRietz2011, Wakhle2014}. The MADs provide a view of the evolution of the fragment mass as a function of the rotation of the dinuclear system, which can be used as a ``clock'' for the time-scale of the reaction~\cite{Toke1985, DuRietz2011}. In fusion-fission, a compound nucleus is formed and undergoes fission, which will occur on a relatively long timescale ($> 10^{-20}-10^{-16}$ s) with no memory of the entrance channel and consequently no mass-angle correlation.  An example of pure fusion-fission is shown in Fig.~1(e) for a $^{12}$C~+~$^{208}$Pb reaction at $E_{\mathrm{c.m.}}/V_{\mathrm{B}}=1.13$. In comparison, quasifission, expected to occur on a shorter timescale ($ < 10^{-20}$ s), will exhibit a mass-angle correlation when the dinuclear system separates within the first half rotation of the system. This correlation is a model-independent indication of quasifission and is clearly present in the MADs from the Cr~+~W reactions [Fig.1(a)-(d)]~\cite{DuRietz2013}.

\par
As the $(N/Z)_{\mathrm{CN}}$ of the system increases, from left to right in Fig. 1, the mass-angle correlations evolve, showing the dependence of quasifission on the neutron-richness of the system. In the more neutron-deficient systems [panels~(a)-(b)] an enhancement in the fission-like fragments near the mass of the projectile-like and target-like fragments is observed at forward and backward angles, respectively. This indicates fast timescale quasifission in which the dinuclear system separates quickly without a large amount of mass transfer. These features are diminished for the neutron-rich reactions [panels~(c)-(d)], where the MADs show an increased production of mass symmetric fragments ($M_{\mathrm{R}}$~=~0.5). This can be seen most clearly in the projected mass ratio distributions [Fig. 1 panel (f) to (i)].  Independent of any fit, there is a clear transition to narrower quasifission mass distribtuions, indicating more mass equilibration resulting from a longer sticking time with increased $(N/Z)_{\mathrm{CN}}$.


\par
To quantify this behavior, a Gaussian function, with mean fixed at $M_{\mathrm{R}} = 0.5$ and width ($\sigma_{\mathrm{exp}}$) allowed to freely vary, was fitted to the fission-like component over the range of $0.35 < M_{\mathrm{R}} < 0.65$.  The experimental distributions were well reproduced by the fitted functions (shown as the solid red lines in Fig.~1(f)-(i))~\cite{Hinde1996, Prokhorova2008, Williams2013, Hinde2008, Itkis2011, Itkis2004}.  The width of the mass distributions expected for pure fusion-fission ($\sigma_{\mathrm{ff}}$) can be estimated from a simple statistical approximation as $\sigma_{\mathrm{ff}} = \sqrt{T/k}$, where $T$ is the scission point temperature and $k$ is the stiffness parameter taken as 0.0048 MeV/u$^{2}$~\cite{Lin2012,Back1996,Back1985}.  These Gaussian functions are shown by the blue dashed lines in panels~(f)-(j).  The mass distribution from the $^{12}$C~+~$^{208}$Pb reaction, where fusion-fission is expected, is well reproduced.  In comparison the mass distributions from the Cr~+~W reactions are severely underestimated by the fusion-fission calculation, consistent with a strong quasifission component.




\begin{figure}
\includegraphics*[scale=0.38]{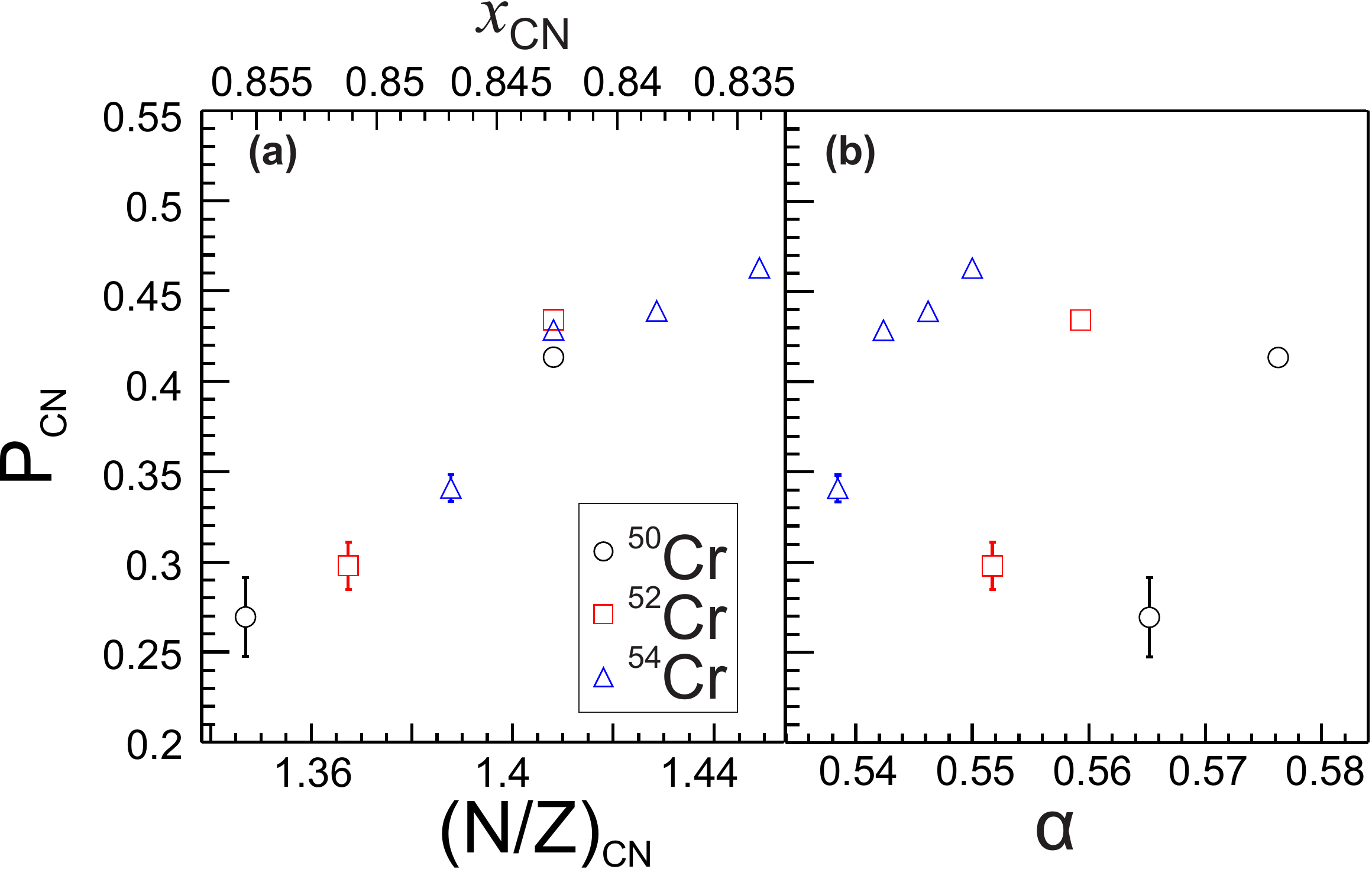}
\caption{Upper-limits of P$_{\mathrm{CN}}$ extracted from the measured mass distributions of each system as a function of (a) $(N/Z)_{\mathrm{CN}}$-bottom axis, $x_{CN}$-top axis, and (b) mass asymmetry, $\alpha $.}
\end{figure}

\par
Upper limits for $P_{\mathrm{CN}}$ from each reaction were estimated from the mass widths using $P_{\mathrm{CN}} = \sigma_{\mathrm{ff}}/\sigma_{\mathrm{exp}}$~\cite{Lin2012} and are shown as a function of the $(N/Z)_{\mathrm{CN}}$ in Fig.~2(a).  The least neutron-rich system has a maximum $P_{\mathrm{CN}}$ of $\sim$25$\%$, while the most neutron-rich system exhibits a significant increase with $P_{\mathrm{CN}} = \sim45\%$.  A smooth linear increase of $P_{\mathrm{CN}}$ with increasing $(N/Z)_{\mathrm{CN}}$ is observed.  These results indicate that quasifission can be suppressed (and fusion enhanced) by increasing $(N/Z)_{\mathrm{CN}}$ and, therefore, strongly promote the use of neutron-rich projectiles in SHE fusion reactions.

\par
An important consideration when characterizing the dependence of quasifission on $(N/Z)_{\mathrm{CN}}$ is to clarify the balance between changing the fissility of the compound nucleus and the mass asymmetry in the entrance channel. A decrease in mass asymmetry is expected to increase quasifission, whilst a decreased fissility has the opposite effect~\cite{Berriman2001,Hinde2002_1,Chizhov2003,Backetal1985,Lin2012,Yanez2013}.  Therefore, when the neutron number of the projectile is increased (e.g. neutron-rich RIB) both the mass-asymmetry and fissility of the system decrease producing opposing effects which must be disentangled in order to understand the $(N/Z)_{\mathrm{CN}}$ dependence of quasifission.  This discrepancy has led to conflicting conclusions from experiments relying heavily on theoretical models which predict either the fissility or mass-asymmetry to be the dominant influence for quasifission~\cite{Back2014, Adamian2012,Adamian2000, Antonenko1993, Antonenko1995, Blocki1986, Adamian2003,Swiatecki1982,Bjornholm1982,Zag2001}. For example, Vinodkumar \emph{et al.}~\cite{Vinod2008} and Sahm \emph{et al.}~\cite{Sahm1985} both reported decreased $P_{\mathrm{CN}}$ with increased $(N/Z)_{\mathrm{CN}}$ for Sn~+~Zr reactions, while Lesko~\emph{et al.}~\cite{Lesko1986} and Liang \emph{et al.}~\cite{Liang2012}, measuring Sn+Ni systems, found the opposite trend.  The measured $P_{\mathrm{CN}}$ upper-limits are shown as a function of fissility (inversely proportional to $(N/Z)_{\mathrm{CN}}$) and mass-asymmetry in panels (a) and (b) of Fig.~2, respectively.  An overall correlation of $P_{\mathrm{CN}}$ with the mass-asymmetry is not present indicating that the strong increase in quasifission with decreasing $\alpha$, obtained in many previous works~\cite{Adamian2000, Antonenko1993, Antonenko1995, Blocki1986, Adamian2003,Berriman2001}, is not seen in these measurements. Thus, the decreased fissility associated with the increasing neutron-richness of the system has a more dominant role in the quasifission process than the change in mass-asymmetry. The $^{50,52,54}$Cr~+~$^{180}$W reactions, shown as the top three points in panel~(b), provide an example where the increased neutron number of the projectile reduces the mass-asymmetry (which would be typically associated with increased quasifission), yet a strong increase in $P_{\mathrm{CN}}$ is observed, owing to the increasing neutron-richness of the projectile.


\begin{figure}
\centering
\includegraphics*[scale=0.3]{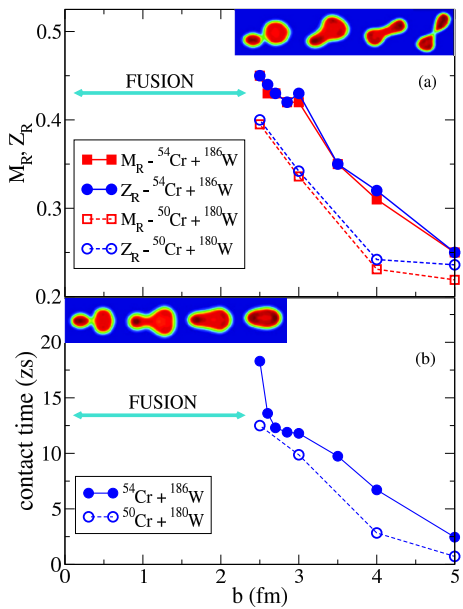}
\caption{TDHF results for the  (a) mass ratio, charge ratio, and (b) contact time as a function of impact parameter for the $^{50}$Cr~$+$~$^{180}$W and $^{54}$Cr~$+$~$^{186}$W systems. In the upper right corners of each panel, characteristic density contours are shown depicting quasifission (top, $^{54}$Cr~$+~^{186}$W at $b=3$~fm) and fusion (bottom, $^{50}$Cr~$+~^{180}$W at $b=0$~fm) events from the TDHF calculations.}
\end{figure}

\begin{figure}
\centering
\includegraphics*[scale=0.3]{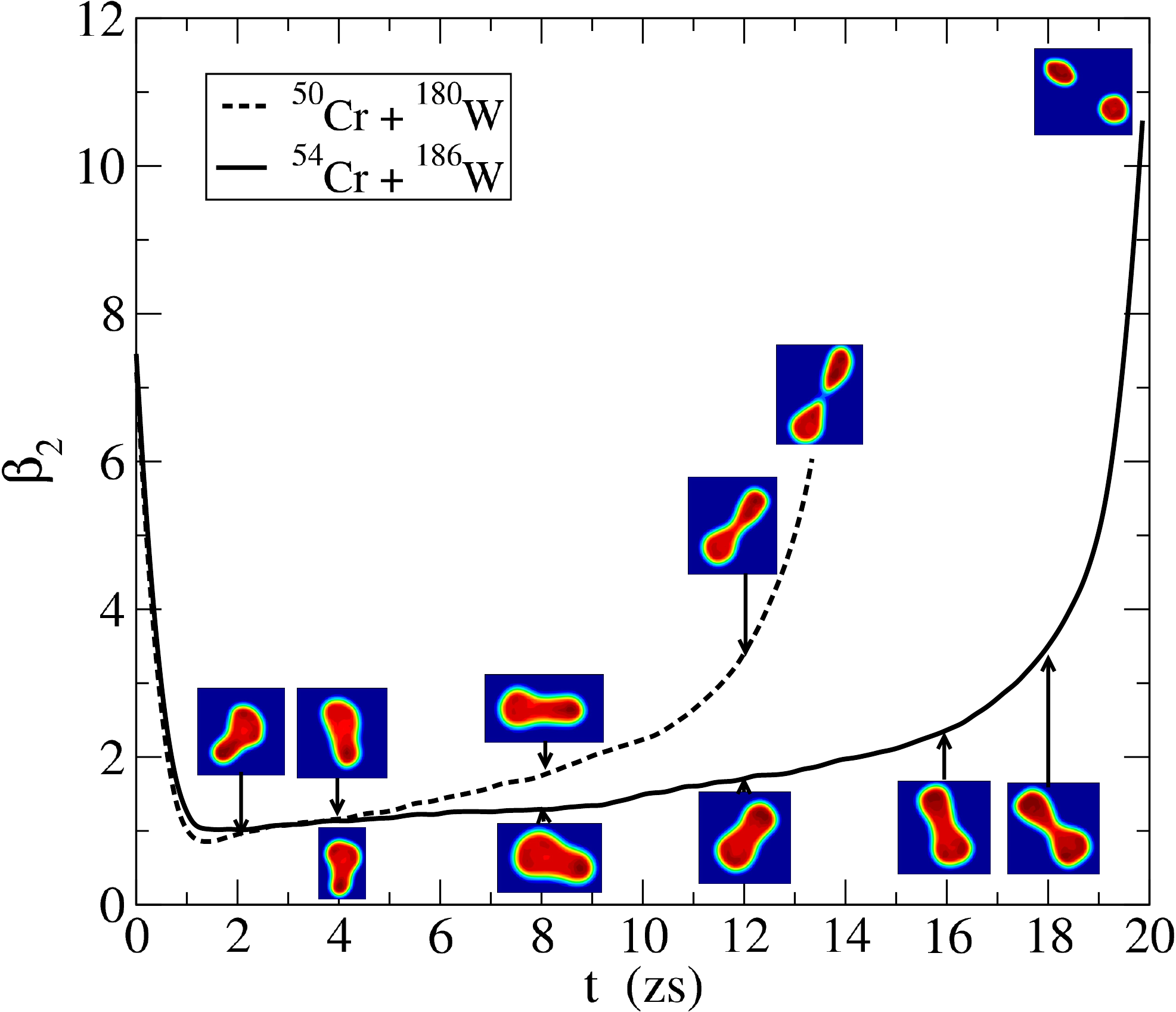}
\caption{Scaled mass quadrupole moment, $\beta_{2}$, as a function of time from the TDHF calculations of the $^{54}$Cr~+~$^{186}$W and $^{50}$Cr~+~$^{180}$W systems at $b=2.5$~fm.  Select density contours are shown along each trajectory.}
\end{figure}

\par
Recently, the TDHF theory has been proposed as a framework to provide a microscopic description of the quasifission process~\cite{Wakhle2014, SimenelEPJA2012, Oberacker2014,Simenel2014}. The $^{50}$Cr~+~$^{180}$W (most neutron-deficient) and $^{54}$Cr~+~$^{186}$W (most neutron-rich) reactions were simulated within the TDHF approach~\cite{Umar2006TDHF}. The evolution of the many-body system was examined as a function of time for different impact parameters $b$.  Both quasifission and fusion (defined here as contact times exceeding 35 zs) were observed. Examples of the density contours from the quasifission and fusion reactions are shown at the top of Fig.~3(a) and (b), respectively. In the quasifission reaction [panel~(a)], it is clear that the projectile and target form a rotating dinuclear system that eventually reaches a scission point. In comparison, a compact mononucleus [panel~3(b)] is formed with no indication that the system will re-separate in the events classified as fusion. Fusion occurs primarily from collisions with the side of the W nucleus, which lead to more compact configurations~\cite{Hinde1995, Hinde1996}. In order to investigate the competition between quasifission and fusion, we performed calculations where the projectile collides with the side of the target.

\par
The evolution of the mass ratio $M_{\mathrm{R}}$, charge ratio $Z_{\mathrm{R}}$, and contact time with impact parameter are shown in Fig.~3. Extensive mass/charge transfer towards symmetry ($M_{\mathrm{R}}=0.5$) and long contact times ($>10$~zs)~\cite{Toke1985, DuRietz2013} are observed for events with $b=2.5-3.0$~fm, characteristic of quasifission reactions.  A clear $(N/Z)_{\mathrm{CN}}$ dependence is observed in the TDHF calculations, with the neutron-rich system exhibiting an increase in the mass/charge transfer and contact time of the quasifission process relative to the neutron-deficient system.  The difference in the evolution of the two reactions is depicted in Fig.4, which shows the scaled mass quadrupole moment, $\beta_2(t)=(4\pi/3)\frac{Q_{20}(t)}{AR_0^2}$, with $R_0=1.2A^{1/3}$. The instantaneous mass quadrupole moment is calculated by diagonalizing the mass quadrupole tensor and picking the eigenvalue corresponding to the symmetry axis. Some instantaneous density profiles are also shown on the figure.  While the initial stages of the reaction are similar, the rate of elongation of the dinuclear system is much faster for the neutron-deficient system, leading to a shorter contact time.  These observations show that the neutron-rich system remains in a compact configuration much longer, which is expected to lead to an increased $P_{\mathrm{CN}}$ value.  The TDHF results are also shown on the MADs in Fig.~1 and display the mass-angle correlations expected from quasifission reactions.


\par
In summary, systematic mass-angle distribution measurements for Cr~+~W reactions were performed to isolate the dependence of quasifission on neutron-richness.  The characteristics of quasifission show a strong dependence on the $N/Z$ of the compound system.  The MADs indicate an increase in the timescale and a decrease in the strength of the quasifission channel for the more neutron-rich systems.  These results demonstrate that the decrease in fissility gained from increasing the neutron content of the projectile outweighs the associated decrease in the mass-asymmetry of the system.   These conclusions are in agreement with microscopic TDHF calculations, which show increased interaction times and fusion cross sections for the most neutron-rich system relative to the most neutron-deficient system. These outcomes strongly support continued exploration of TDHF as a microscopic tool for studying low-energy heavy-ion reaction dynamics and encourage the use of the increasingly neutron-rich projectiles in future SHE reactions, including the consideration of using exotic neutron-rich radioactive ion beams.

\emph{Acknowledgments.}  The authors are grateful for the high quality beams provided by the staff at the ANU accelerator facility.  This work
is supported by the National Science Foundation under Grants No. PHY-1102511 and No. IIA-1341088, by the U.S. Department of Energy under Grant No.
DE-FG02-96ER40975 with Vanderbilt University, and the Australian Research Council Grants DP110102858, DP140101337, FL110100098, DP130101569, FT120100760, and DE140100784.  This material is based upon work supported by the Department of Energy National Nuclear Security Administration under Award Number DE-NA0000979. This material is based upon work supported by the U.S. Department of Energy, Office of Science, Office of Nuclear Physics, under Contract No. DE-AC02-06CH11357. This research used resources of ANL's ATLAS facility, which is a DOE Office of Science User Facility.



%

\end{document}